\documentclass[11pt]{article}
\usepackage{epsfig}

\newcommand{\bm}[1]{\mbox{\boldmath$#1$}}
\def\mvec#1{{\bm{#1}}}  

\begin{document}

\title{From Observations to Hypotheses \\
{\Large Probabilistic Reasoning Versus Falsificationism 
and its   Statistical Variations}\\ \mbox{}}
\author{G. D'Agostini \\
{\small Universit\`a ``La Sapienza'' and INFN, Rome, Italy} \\
{\small (giulio.dagostini@roma1.infn.it, 
www.roma1.infn.it/\,$\tilde{ }$\,dagos) } \\ \mbox{}
}

\date{}

\maketitle

\begin{abstract}
Testing hypotheses is an issue of primary importance 
in the scientific research, as well as in many other human activities.
Much clarification about it
can be achieved if the process of learning 
from data is framed in a stochastic model of causes and effects.
Formulated with Poincar\'e's words, 
the {\it ``essential problem of the experimental method''}
becomes then solving a {\it ``problem in the probability of causes''}, 
i.e. ranking  the several hypotheses,
that might be responsible for the observations, in credibility. 
This probabilistic approach to the problem 
(nowadays known as the Bayesian approach) differs from the
standard (i.e. frequentistic) statistical methods of hypothesis tests. 
The latter methods might be seen as practical
 attempts of implementing 
the ideal of falsificationism, that can  itself be viewed 
as an extension of the proof by contradiction 
of the classical logic to the experimental method. 
Some criticisms concerning conceptual as well as practical
aspects of na\"\i ve falsificationism and conventional,
frequentistic hypothesis tests are presented, 
and the alternative, probabilistic approach is 
outlined.
\end{abstract}

\mbox{} \\
\vspace{1.5cm}
\mbox{} \\
\noindent
Invited talk at the 2004 Vulcano Workshop on {\sl Frontier Objects in Astrophysics and Particle Physics}, Vulcano (Italy) May 24-29, 2004. 
\newpage
\section{Inference, forecasting and related uncertainty}
\begin{figure}
\centering\epsfig{file=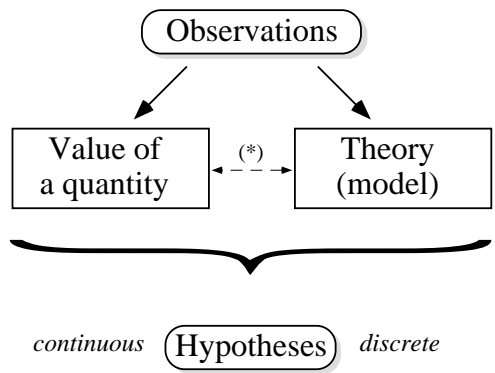,clip=,width=6.5cm}
\caption{From observations to hypotheses. $^{(*)}$\,The link 
between value of a quantity and theory is a reminder 
that sometimes a quantity has meaning only 
within a given theory or model~\cite{BR}.}
\label{fig:obs_hyp}
\end{figure}

The intellectual process of learning from observations can be
sketched as illustrated in figure \ref{fig:obs_hyp}.
From experimental data we wish to `determine' the value of 
some physical quantities, 
or to establish which theory describes `at best'
the observed phenomena. Although these two
tasks are usually 
seen as separate issues, and analyzed with different
mathematical tools, they  
can be viewed as two subclasses of the same process: 
{\it inferring hypotheses from observations}.
What differs between the two kinds of inference
is the number of hypotheses that enters the game:
a discrete, usually small number when dealing with 
{\it theory comparison}; 
a large, virtually infinite number  
when {\it inferring the value of physical quantities}.

\begin{figure}
\centering\epsfig{file=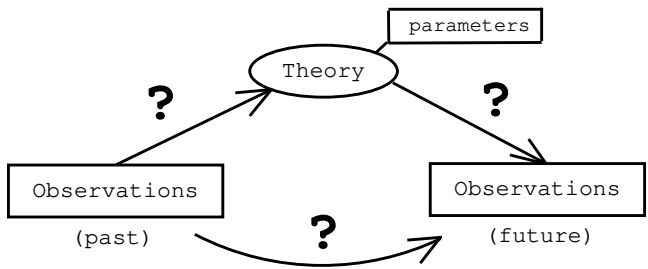,clip=,width=8.5cm}
\caption{Theory (and the 
value of its parameters) acting as a {\it link} between
past and future.}
\label{fig:past-future}
\end{figure}

\begin{figure}
\centering\epsfig{file=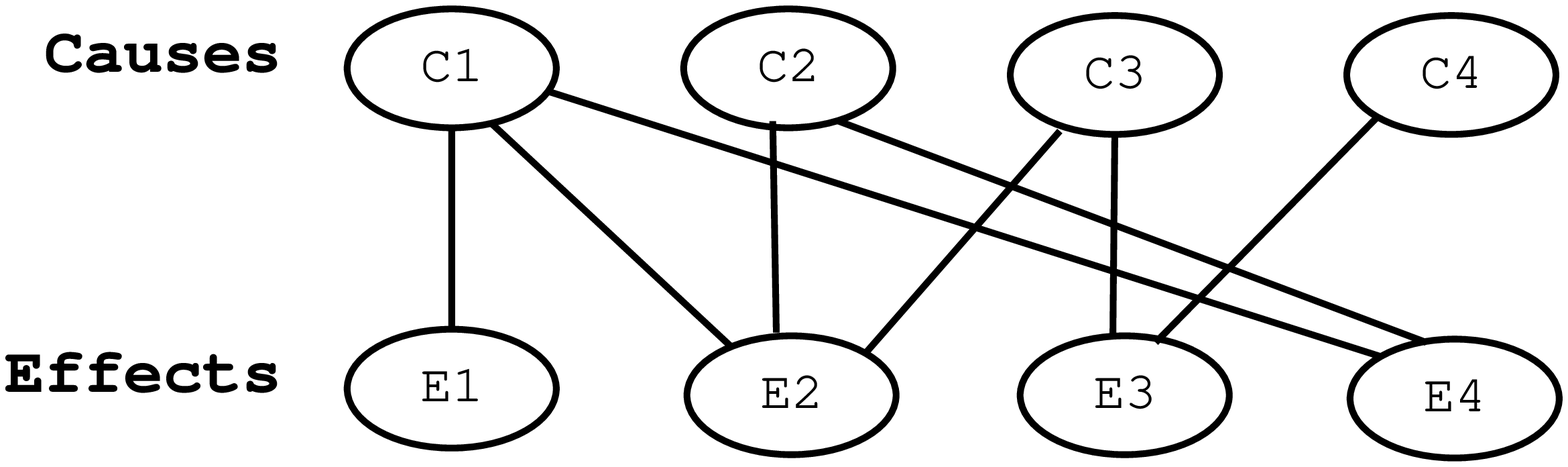,clip=,width=9.5cm}
\caption{Causal links (top-down) and inferential links (down-up).}
\label{fig:cause-effetti}
\end{figure}

In general, given some data ({\it past observations}), 
we wish to:
\begin{itemize}
\item
 select a theory and determine its parameters
with the aim to  describe and `understand' the physical world; 
\item
predict {\it future observations} (that, once they are recorded, 
they join the set of past observations to corroborate or diminish
our confidence on each theory and its parameters). 
\end{itemize}
The process of learning from data and predicting new 
observations is characterized by {\it uncertainty}
 (see figure \ref{fig:past-future}). 
Uncertainty in
going from past observations to the theory and its parameters.
Uncertainty in predicting precise observations from the theory.
And, as a consequence, uncertainty in predicting 
future observations
from past observations. 
Rephrasing the 
hypothesis-observation scheme 
in terms of {\it causes} and {\it effects},
we can realize that  
the very source of uncertainty is  
due to the not biunivocal relationship  between causes and effects, 
as sketched in  figure 
\ref{fig:cause-effetti}. 
The fact that {\it identical} causes 
--- identical according to our knowledge 
--- might produce different effects can be due to
{\it internal} (intrinsic) probabilistic aspects of the theory, as well as to 
our lack of knowledge about the exact set of 
causes.\footnote{One might object that if the same cause yields 
different effects in different trials, then other concauses must exist, 
responsible for the differentiation of the effects. 
This point of view leads e.g. to the `hidden variables' interpretation
of quantum mechanics (`\`a la Einstein'). 
I have no intention to try to solve, or even to touch all philosophical 
questions related to causation (for a modern and fruitful approach, 
see Ref.~\cite{Causality} and references therein)
and of the fundamental aspects of
quantum mechanics.
The approach followed here 
is very pragmatic and the concept of causation is, to say, 
a {\it weak} one,
that perhaps could be better called {\it conditionalism}: 
``whenever I am sure of {\it this}, then I am 
also somehow confident that {\it that} will occur''. 
The degree of confidence on the occurrence of {\it that} 
might rise from past experience, just from reasoning, 
or from both.
It is not really relevant whether
{\it this} is the cause of {\it that} in a classical sense,
or {\it this} and {\it that} are both due to other `true causes' 
and we only perceive a correlation between  {\it this} and {\it that}.}
(Experimental errors are one of the components of the {\it external}
probabilistic behavior of the observations.)
However, there is no practical difference between the two situations,
as far as 
 the probabilistic behavior of the result is concerned 
(i.e. in the status of our mind concerning the possible outcomes of the experiment), 
and hence to the probabilistic character of inference.

Given this  cause-effect scheme, 
having observed  an effect, we cannot be sure about its cause.
(This is what 
 happens to effects $E_2$, $E_3$ and $E_4$ 
of figure \ref{fig:cause-effetti} ---
effect $E_1$, that can only be 
due to cause $C_1$, 
has to be considered an exception, at least in the inferential problems
scientists typically meet.)
\newpage
\begin{description}
\item[Example 1.]
As a simple example, think about the effect 
identified by the number $x=3$ 
resulting by one of the following random generators 
chosen at random:
$H_1$ = ``a Gaussian generator with $\mu=0$ and $\sigma=1$'';
$H_2$ = ``a Gaussian generator with $\mu=3$ and $\sigma=5$'';
$H_3$ = ``an exponential generator with $\tau=2$''
($\tau$ stands for the expected value of the exponential distribution;
$\mu$ and $\sigma$ are the usual parameters of the Gaussian distribution). 
Our problem, stated in intuitive terms, is to find out which 
hypothesis might have caused $x=3$: $H_1$, 
$H_2$ or $H_3$? Note that none of the hypotheses of this example 
can be excluded and, therefore, 
there is no way to reach a boolean conclusion. We can 
only state, somehow, our {\it rational preferences}, based
on the experimental result and 
our best knowledge of the behavior of each {\it model}.
\end{description}
The human mind is used to live --- and survive ---
 in conditions of 
uncertainty and has developed mental categories to handle it.
Therefore, 
although we are in a constant status of  uncertainty 
about many events which might or might not occur,
we can be ``more or less {\it sure} --- or {\it confident} ---
 on something than on something else''. 
In other words, 
``we consider something more or less {\it probable} 
(or {\it likely})'', or
``we {\it believe} something  more or less than
 something else''. We can use similar expressions, 
all referring to the intuitive idea of {\it probability}.

The status of uncertainty 
 does not prevent us from 
doing Science. Indeed, said with Feynman's words,
 {\it ``it is scientific 
only to say what is more likely and what is less 
likely''}\,\cite{Feynman}. 
Therefore, it becomes crucial to 
learn how to deal quantitatively with probabilities of causes,
because the {\it ``problem}(s) {\it  in the probability of causes \ldots
may be said to be the essential problem}(s) {\it  of
the experimental method''} (Poincar\'e\,\cite{Poincare}).

However, and unfortunately, it is a matter of fact that 
nowadays most scientists are 
incapable to reason correctly about probabilities of causes, 
probabilities of hypotheses, probabilities 
of values of a quantities,
and so on. This lack of expertise is due to 
the fact that we have been educated and trained
with a statistical theory in which the very concept of probability
of hypotheses is absent, although we naturally tend to think and 
express ourselves in such terms. In other words, the common prejudice
is that probability {\it is} the long-term relative frequency,
but, on the other hand, probabilistic statements about hypotheses
(or statements implying,
anyway, a probabilistic meaning) are constantly made by
the same persons, statements that are 
irreconcilable with their definition of 
probability~\cite{Maxent98}.
The result of this
mismatch between natural thinking and cultural 
over-structure produces mistakes in scientific judgment, 
as discussed e.g. in Refs.~\cite{BR,Maxent98}.

Another prejudice, rather common among scientists, is that, when they
deal with hypotheses, `they think they reason' according to the 
falsificationist scheme: hence, the hypotheses tests of conventional 
statistics are approached with a genuine intent of proving/falsifying 
something. For this reason we need to shortly review these concepts,
in order to show the reasons why they are less satisfactory than 
we might na\"\i vely think. (The reader is assumed to be familiar with
the concepts of hypothesis tests, though at an elementary level 
 --- null hypothesis, one and two tail tests, 
acceptance/rejection, significance, type 1 and type 2 errors, an so on.)

\section{Falsificationism and its statistical variations}
\label{falsificationism}
The essence of the so called {\it falsificationism}
is that a theory should yield {\it verifiable} predictions,
i.e. predictions that can be checked
to be {\it true} or {\it false}. If an effect is observed 
that contradicts the theory, the theory is ruled out, i.e. it is
{\it falsified}. 
Though this scheme is 
certainly appealing, and most scientists are convinced that this
is the way Science proceeds,\footnote{Those who believe that 
scientists are really `falsificationist' can find enlighting 
the following  famous Einstein's quote: 
{\it ``If you want to find out anything from 
the theoretical physicists about the methods they use, 
I advise you to stick closely to one principle: 
don't listen to their words, fix your attention 
on their deeds.}''\cite{Einstein}. We shall come to this point
in the conclusions.
}
 it is easy to 
realize that this scheme
 is a bit na\"\i ve, when one tries to apply it literally,
as we shall see in a while. Before doing that, 
it is important 
to recognize that falsificationism is nothing but an  
extension of the classical {\it proof by contradiction}
to the experimental method.

The proof by contradiction of standard dialectics and mathematics 
consists in assuming true a hypothesis and in looking 
for (at least) one of its logical consequences
that is manifestly false.
If a false consequence exists, 
then the hypothesis under test is considered false
and its opposite true (in the sequel $\overline H$
will indicate
the hypothesis opposite to $H$, i.e. $\overline H$ is
true if  $H$ false, and vice versa).
Indeed, there is no doubt that if we observe an effect that 
 is {\it impossible} within a theory, this theory 
has to be ruled out. But the {\it strict} application of the 
falsificationist criterion is not maintainable in the 
scientific practice for several reasons.
\begin{enumerate}
\item
What should we do of all theories which have not been falsified yet?
Should we consider them all at the same level, parked  
in a kind of {\it Limbo}? This approach is not very   
effective. Which experiment should we perform next? The natural
development of Science shows that new investigations are made
in the direction that seems  mostly {\it credible}
(and fruitful) at a given moment, a behaviour
 often {\it ``179 degrees or so out of phase from Popper's
idea that we make progress by falsificating theories''}~\cite{Wilczek}.
\item
If the predictions of a theory are characterized by the 
internal or external probabilistic behavior discussed above, 
how can we ever think of
 falsifying such a  theory, speaking rigorously?
For instance, there is no way to falsify hypothesis $H_1$ 
of Example 1, because any real number is compatible 
with any Gaussian. 
For the same reason, falsificationism cannot be used to make 
an inference about the value of a physical quantity
(for a Gaussian response of the detector,
no value of $\mu$ can be falsified whatever we observe, 
and, unfortunately, 
 falsificationism does not tell how to classify  non-falsified values
in credibility). 
\end{enumerate}
An extension of strict falsificationism 
is offered by the 
{\it statistical test} methods developed by statisticians.
Indeed, the latter methods might be seen as attempts of
implementing in practice the falsificationism principle. 
It is therefore important to understand the `little' variations of 
the statistical 
tests with respect to the proof of contradiction 
(and hence to strict falsificationism).
\begin{enumerate}
\item[a)]
The impossible consequence is replaced by an 
{\it improbable consequence}.
If this improbable consequence 
occurs, then the hypothesis is rejected, otherwise it is accepted.  
The implicit argument on the basis of the 
hypothesis test approach of
 conventional statistics
is:  ``if $E$ is {\it practically impossible} given $H$,
then $H$ is considered {\it practically false} given 
the observation $E$.'' 
But this probability inversion --- initially qualitative, 
but then turned erroneously quantitative by most practitioners,
 attributing to `$H$ given $E$' the 
same probability of `$E$ given $H$' --- is not logically 
justified and it is not difficult to show that it 
yields misleading conclusions.  
Let us see some simple examples.
\begin{description}
\item[Example 2] 
Considering \underline{only} hypothesis $H_1$ of Example 1
and taking $E=$ ``$4\le x \le 5$'', we can
calculate the probability of obtaining $E$ from $H_1$:
 $P(E\,|\,H_1)=3\times10^{-5}$. This probability is rather small, but, 
once $E$ has occurred, we cannot state that 
``$E$ has little probability to come from $H_1$'',  or that
``$H_1$ has little probability to have caused $E$'':
 $E$ is \underline{certainly} due to $H_1$!
\item[Example 3]
``I play honestly at lotto, betting on a rare combination'' ($=H$)
and ``win'' ($=E$). You cannot say that since $E$ is 
`practically impossible' given $H$, 
then hypothesis $H$ has to be `practically  excluded', after you have got
the information that I have won [such a conclusion would imply that
it is `practically true' that ``I have cheated'' ($=\overline H$)]. 
\item[Example 4]
An AIDS test to detect HIV infection is perfect to tag 
HIV infected people as `positive' (=\,Pos), 
i.e. $P(\mbox{Pos}\,|\,\mbox{HIV})=1$, but it can sometimes err, and 
classify healthy persons 
($=\overline{\mbox{HIV}}$) as positive, 
although with low probability, e.g.
$P(\mbox{Pos}\,|\,\overline{\mbox{HIV}})=0.2\%$. 
An Italian citizen is chosen
{\it at random} to undergo such a test and he/she is tagged positive.
We cannot claim that ``since it was practically impossible that a 
healthy person resulted positive, then this person is practically 
infected'', or, quantitatively, 
``there is only 0.2\% probability that this 
person is not infected''. 
\end{description}
We shall see later how to solve these kind of problems correctly. 
For the moment the important message is that {\it it is not
correct to replace `improbable' in logical methods that
speak about `impossible'} (and to use then the reasoning 
to perform `probabilistic inversions'): impossible and improbable differ
in quality, not just in quantity! 
\item[b)]
In many cases the number of effects due to a hypothesis is so large
that {\it each effect is `practically 
impossible'}.\footnote{In the hypothetical experiment of
one million tosses of a hypothetical `regular coin'
(easily realized by a little simulation) 
the result of  500\,000 heads
represents an `extraordinary event' ($8\times 10^{-4}$  probability), 
as `extraordinary' are all other possible outcomes!} 
Even those who trust the reasoning based on the 
small probability of effects to falsify hypotheses 
have to realize that the reasoning fails in these cases, 
because every observation
can be used as an evidence against the hypothesis 
to be tested. 
 Statisticians have then
worked out methods in which 
the observed effect 
is replaced by two {\it ensembles of effects}, 
one of high chance and another of low chance. 
The reasoning based on the `practically impossible'  effect 
is then extended to
the latter ensemble.
This is the essence of all tests 
tests based on ``p-values''\cite{p-values}
 (what physicists know as ``probability\break\newpage
of tails'' upon which $\chi^2$ and other famous tests are based).
Logically,\footnote{The fact that in practice these methods `often work'
is a different story, as discussed in Sec. 10.8 of Ref.~\cite{BR}.}
 the situation gets worse, because
conclusions do not depend anymore on what has been 
observed,
but also 
on effects that have not been observed\footnote{In other words, 
the reasoning based on p-values~\cite{p-values} 
constantly violates the so called {\it likelihood principle}, 
apart from exceptions due to 
numerical coincidences. In fact, making the 
simple example of a single-tail 
test based on a variable that is indeed observed, the 
conclusion about acceptance or rejection is made on the basis of
$\int_{x_{obs}}^\infty\!f(x\,|\,\mvec\theta)\,\mbox{d}x$,
where $\mvec\theta$ are the model parameters. 
But this integral is rarely simply proportional
 to the likelihood $f(x_{obs}\,|\,\mvec\theta)$, i.e. integral and likelihood
do not differ 
by just a constant factor not depending  on $\mvec\theta$.
I would like to make clear that I dislike
un-needed principles,
including the likelihood one, 
and the maximum likelihood one above all.
The reason why I refer here to the likelihood principle in my argumentation
is that, generally, frequentists consider this principle with 
some respect, but their methods usually violate it~\cite{Zech}. 
Instead, in the probabilistic approach illustrated in the sequel, 
this 'principle' stems automatically from the theory.}
(see e.g. Ref.~\cite{BB}).
\item[c)] Apart from the simple case of just one observation,
the data are summarized by a `test variable' (e.g. $\chi^2$), 
function of the data,  
and the reasoning discussed above is applied 
to the test variable. 
This introduces an additional, arbitrary ingredient 
to this already logically tottering 
construction.\footnote{In statistics the variables that
summarize all the information sufficient for the inference
are called a {\it sufficient statistics} (classical 
examples are the sample average and standard deviation to infer
$\mu$ and $\sigma$ of a Gaussian distribution). 
However, I do not know of test variables that 
can be considered sufficient for hypothesis tests.}
\item[d)] Even in simple problems, that could be formulated in 
terms of a single quantity, 
given the empirical information
there might be ambiguity 
about which quantity plays the role of the random variable upon which
the p-value has to be 
calculated.\footnote{Imagine you have to decide if the extraction of
$n$ white balls in 
$N$ trials can be considered in agreement with the hypothesis
that the box contains a given percentage $p$ of 
white balls. You might think that 
you are dealing with a binomial problem, 
in which $n$ plays the role of random variable,
calculate the p-value and draw your conclusions.
But you might get the information that 
the person who made the extraction had decided to 
go on until he/she reached
$n$ white balls. In this case 
the random variable is $N$, the problem is modeled 
by a Pascal distribution (or, alternatively, by a negative
binomial in which the role of  random variable is played
by the number $N-n$ of non-white balls) 
and the evaluation of the p-value differs 
from the previous one. This problem is known as the {\it stopping rule} problem.
It can be proved that the likelihood calculated from the two reasonings
differ only by a constant factor, and hence the likelihood principle 
tells that the two reasonings should lead to identical inferential conclusions
about the unknown percentage of white balls. 
}
\end{enumerate}
Anyhow, apart from questions that might seem subtle philosophical
quibbles, conventional tests lead to several practical problems. 
\begin{itemize}
\item
In my opinion the most serious problem is the fact that p-values 
are constantly used in scientific conclusions 
 as if they were the probability that the hypothesis 
under test is true (for example people report a p-value of 0.0003
as ``the hypothesis is excluded at 99.97\% C.L.'', as if they were 
99.97\% confident that the hypothesis to test is false). 
The consequence of this misunderstanding is very serious, and it is
essentially responsible for  all claims of fake discoveries 
in the past decades
(see some examples in Sec. 1.9 of Ref.~\cite{BR}.)
\item
Statistical tests are not based on first principles of any kind. 
Hundreds of statistical tests have been contrived and their choice 
is basically arbitrary.  I have experienced that
discussions in experimental teams about which
test to use and how to use it are not deeper than discussions
in pubs among soccer fans (Italian readers might think at the
`Processo di Biscardi' talk show, quite often also in the tones).
\item
There is sometimes a tendency to look for the test that
gives the desired result. Personally, I find that 
{\it the fancier
the name of the test is, the less believable the claim is},
because I am pretty sure that other,
more common tests were discarded
because `they were not appropriate', an expression to be 
often interpreted
as ``the other tests did not support what 
the experimentalist 
 wanted the data to 
prove'' 
 (and I could report of people
that, frustrated by the `bad results' obtained with frequentistic tests,
 contacted me hoping for a Bayesian miracle --- 
they got regularly 
disappointed because, `unfortunately',  
Bayesian methods, consciously applied,
tend not to feed vain illusions).
\item
Standard statistical methods, essentially a contradictory collection of 
{\it ad-hoc-eries}, induce scientists, and physicists in particular,
to think that `statistics' is something `not serious',
thus encouraging
`creative' behaviors.\footnote{Just in this workshop I have 
met yet another invention~\cite{SK}: 
Given three model fits to data with
40 degrees of freedom and the three 
resulting $\chi^2$ of 37.9, 49.1 and 52.4 for
models $M_1$, $M_3$ and $M_3$, the common frequentistic wisdom 
says the three models are about equivalent in describing the data,
because the expected $\chi^2$ is $40\pm 9$, or that none of the models
can be ruled out because all p-values  
(0.56, 0.15 and 0.091, respectively) are above the usual 
critical level of significance.
Nevertheless, SuperKamiokande claims that models $M_2$ and $M_3$ are 
`disfavored' at 3.3 and 3.8 $\sigma$'s, respectively! ($1\times 10^{-3}$
and $1.4\times 10^{-4}$ probability.) It seems 
the result has been achieved using
inopportunely a technique of parametric inference. Imagine a minimum
$\chi^2$ fit of the parameter $\theta$ for which the data
give a minimum $\chi^2$
of 37.9 at $\theta=\theta_1$, while $\chi^2(\theta_2)=49.1$ and 
$\chi^2(\theta_3)=52.4$ (and the $\chi^2$ curve is parabolic). 
It follows that $\theta_2$ and $\theta_3$ are, respectively,  
$\sqrt{49.1-37.9}\,\sigma\mbox{'s} =3.3\,\sigma$'s and 
$\sqrt{52.4-37.9}\,\sigma\mbox{'s}=3.8\,\sigma$'s far 
from $\theta_1$. The probability 
that $\theta$ differs from $\theta_1$ by more than 
 $|\theta_2-\theta_1|$ and  $|\theta_3-\theta_1|$ is then
$P(|\theta-\theta_1|>|\theta_2-\theta_1|) = 1\times 10^{-3}$ 
and  
$P(|\theta-\theta_1|>|\theta_3-\theta_1|) = 1.4\times 10^{-4}$,
 respectively.
But this is quite a different problem!}  
\end{itemize}

\section{Forward to the past: probabilistic reasoning}\label{probabilistic}
The dominant school in statistics since the beginning of last
century is based on a quite unnatural approach to probability,
in contrast to that of the founding fathers (Poisson,
Bernoulli, Bayes, Laplace, Gauss, etc.). In this dominant approach 
({\it frequentism}) there is no room for the concept of probability
of causes,
probability of hypotheses, probability 
of the values of physical quantities, 
and so on. 
Problems in the probability of the causes 
(``{\it the essential problem of the experimental method}''!\,\cite{Poincare})
have been replaced by the machinery of the hypothesis tests.
But people think naturally in terms of probability of causes,
and the mismatch between natural thinking and standard
education in statistics leads to the troubles
discussed above. 

I think that the way out is simply to {\it go back to the past}.
In our time of rushed progress an invitation to
go back to century old ideas seems at least odd (imagine 
a similar proposal regarding physics, chemistry or biology!).
I admit it, but I do think it is the proper way to follow. 
This doesn't mean we have to drop everything done in probability and statistics 
in between. Most mathematical work can be easily recovered.
In particular, we can benefit of theoretical clarifications and 
progresses in probability theory of the past century. 
We also take great advantage of the boost of computational capability
occurred very recently, from which both  symbolic and numeric
methods have enormously benefitted.
(In fact, many frequentistic ideas had their {\it raison d'\,\^etre}
in the computational barrier that the original probabilistic
approach met. Many simplified -- though often 
simplistic -- methods were then  proposed
to make the live of practitioners easier.
But nowadays computation cannot be considered any longer an excuse.) 

In summary, the proposed way out can be summarized
in an invitation to {\it use probability theory  consistently}.
But before you do it, you need to 
review the definition of probability, otherwise it is simply
impossible to use all the power of the theory. 
In the advised approach probability
quantifies how much we believe in something, i.e. we recover its 
intuitive idea. Once this is done, we can 
essentially use the formal probability theory based on Kolmogorov axioms
(which can indeed be derived, and with a better
awareness about their meaning, 
from more general principles! -- but 
I shall not enter this issue here). 

This `new'
approach is called {\it Bayesian} because of the central 
role played by Bayes theorem in learning from experimental data. 
The theorem teaches how the probability of each hypothesis $H_i$ 
has to be updated in the light of the new observation $E$:
\vspace{-0.5mm}
\begin{eqnarray}
P(H_i\,|\,E,I) &=& \frac{P(E\,|\,H_i,I)\cdot P(H_i\,|\,I)}
                        {P(E\,|\,I)}\,.
\label{eq:Bayes0}
\end{eqnarray}
\vspace{-0.5mm}
$I$ stands for a background condition, or status of 
information, under which the inference is made. 
A  more frequent Bayes' formula in text books, valid if the hypotheses 
are exhaustive and mutually exclusive, is 
\vspace{-0.5mm}
\begin{eqnarray}
P(H_i\,|\,E,I) &=& \frac{P(E\,|\,H_i,I)\cdot P(H_i\,|\,I)}
                        {\sum_iP(E\,|\,H_i,I)\cdot P(H_i\,|\,I)}\,.
\label{eq:Bayes}
\end{eqnarray}
\vspace{-0.5mm}
The denominator in the right hand 
side of (\ref{eq:Bayes})
is just a normalization factor and, as such,
it can be neglected. Moreover it is possible to
show that a similar structure holds for probability density functions
(p.d.f.) if a continuous variable is considered ($\mu$ stands here
for a generic `true value', associated to a parameter of a model). 
Calling `data' the overall effect $E$,
we get the following formulae on which inference is to be ground:
\vspace{-0.5mm}
\begin{eqnarray}
P(H_i\,|\,\mbox{data},I) &\propto& 
P(\mbox{data}\,|\,H_i,I)\cdot P(H_i\,|\,I) 
\label{eq:Bayes_1}\\
f(\mu\,|\,\mbox{data},I) &\propto& f(\mbox{data}\,|\,\mu,I)
                          \cdot f(\mu\,|\,I)\,,\label{eq:Bayes_2}
\end{eqnarray}
\vspace{-0.5mm}
the first formula used in probabilistic comparison of hypotheses,
the second (mainly) in {\it parametric inference}.
 In both cases
we have the same structure:
\vspace{-0.5mm}
\begin{eqnarray}
\mbox{\bf posterior} &\propto& \mbox{\bf likelihood} \times 
\mbox{\bf prior}\,,
\label{eq:bayes_structure}
\end{eqnarray}
\vspace{-0.5mm}
where `posterior' and `prior'
refer to our belief on that hypothesis, i.e. 
taking or not taking into account the `data' on which the
present inference is based. The likelihood, 
that is ``how much we believe 
that the hypothesis
can produce the data'' (not to be confused with 
``how much we believe that the data
come from the hypothesis''!), models the stochastic flow that
leads from the hypothesis to the observations, including the 
best modeling of the detector response. The structure
of (\ref{eq:bayes_structure}) shows us that the inference based on Bayes
theorem satisfies automatically the {\it likelihood principle}
(likelihoods that differ by constant factors lead to the 
same posterior).

The proportionality factors in (\ref{eq:Bayes_1}) and (\ref{eq:Bayes_2}) 
are determined by normalization, if absolute probabilities are needed.
Otherwise we can just put our attention on probability ratios:
\begin{eqnarray}
\frac{P(H_i\,|\,\mbox{data},I)}
     {P(H_j\,|\,\mbox{data},I)} &=& 
\frac{P(\mbox{data}\,|\,H_i,I}{P(\mbox{data}\,|\,H_j,I)}
\cdot\frac{P(H_i\,|\,I)}{P(H_j\,|\,I)} \label{eq:Bayes_ratios} \\
\mbox{i.e.}\hspace{4cm} && \mbox{    } 
\nonumber \\
\mbox{\bf posterior odds} &\propto& \mbox{\bf Bayes factor} \times 
\mbox{\bf prior odds}\,:
\end{eqnarray}
{\it odds} are updated by data via the ratio of the 
likelihoods, called {\it Bayes factor}.

There are some well known psychological (indeed 
cultural and even ideological) resistances to this approach 
due to the presence of the priors in the theory. 
Some remarks are therefore in order:
\begin{itemize}
\item   
priors are unescapable, if we are interested in 
`probabilities of causes' (stated differently, 
there is no other way to relate consistently  
probabilities of causes 
to probabilities of effects avoiding Bayes theorem);
\item
therefore, you should mistrust methods that
pretend to provide `levels of confidence'
(in the sense of {\it how much you are confident})
independently from priors (arbitrariness is often sold for objectivity!);
\item
in many `routine' applications the results 
of measurements depend weakly on priors and 
many standard formulae, usually derived from 
maximum likelihood or least square principles, can be promptly 
recovered
{\it under well defined conditions of validity},
\item
but in other cases priors might have a strong 
influence on the conclusions;
\item
if we understand the role and the relevance of the priors, we 
shall be able to provide useful results in 
the different cases (for example, when the priors 
dominate the conclusions and there is no agreement 
about prior knowledge, it is better  
to refrain from providing probabilistic results:
Bayes factors may be considered a convenient way to 
report how the experimental data push toward either
hypothesis; similarly, 
upper/lower ``xx\%\ C.L.'s'' are highly misleading and 
should be simply replaced by {\it sensitivity bounds}\,\cite{BR}).
\end{itemize}
To make some numerical examples, let us solve two of the 
problems met above. 
(In order to simplify the notation  the  
background condition `$I$' is not indicated explicitly 
in the following formulae).
\begin{description}
\item[Solution of the AIDS problem (Example 4)]\mbox{}\\
Applying Eq.~(\ref{eq:Bayes_ratios}) we get
\begin{eqnarray}
\frac{P(\mbox{HIV}\,|\,\mbox{Pos})}{P(\overline{\mbox{HIV}}\,|\,\mbox{Pos})} 
&=& \frac{P(\mbox{Pos}\,|\,\mbox{HIV})}
{P(\mbox{Pos}\,|\,\overline{\mbox{HIV}})} 
\cdot \frac{P(\mbox{HIV})}{P(\overline{\mbox{HIV}})}\,.
\label{eq:ADIS_sol}
\end{eqnarray}
The Bayes factor $P(\mbox{Pos}\,|\,\mbox{HIV})/
P(\mbox{Pos}\,|\,\overline{\mbox{HIV}})$
is equal to 1/0.002 = 500. This is 
how much the information provided by the data `pushes' 
towards the hypothesis `infected' with respect to 
the hypothesis `healthy'. {\it If} the ratio of priors 
were equal to 1 [i.e. $P(\mbox{HIV})=P(\overline{\mbox{HIV}})$\,!], 
we would get final odds of 500, i.e. 
$P(\mbox{HIV}\,|\,\mbox{Pos})=500/501= 99.8\%$. But, fortunately, 
for a randomly chosen  Italian $P(\mbox{HIV})$ is not 50\%. 
Putting some more 
reasonable numbers, that might be 1/600 or 1/700, 
we have final odds of 0.83 or 0.71, corresponding 
to a $P(\mbox{HIV}\,|\,\mbox{Pos})$ of 45\% or 42\%. We understand 
now the source of the mistake done by quite some people in front
of this problem: priors were unreasonable! 
This is a typical situation: using the Bayesian reasoning
it is possible to show the hidden assumptions of
non-Bayesian reasonings, though most users of the latter
methods object, insisting in claiming they ``do not use priors''.
\item[Solution of the three hypothesis problem (Example 1)]\mbox{}\\
\begin{figure}
\centering\epsfig{file=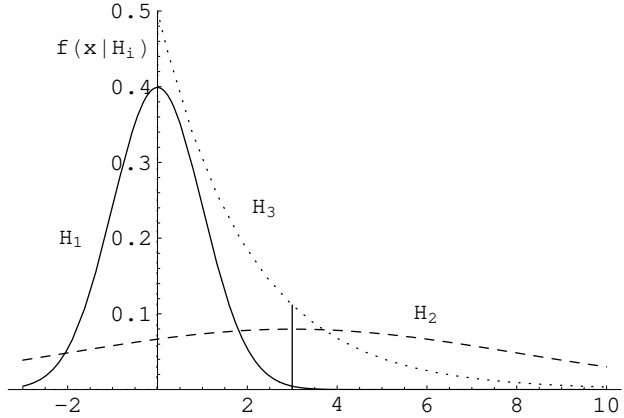,clip=,width=9.0cm}
\caption{Example 1: likelihoods for the three different hypotheses. 
The vertical bar corresponds to the observation $x=3$.}
\label{fig:example1}
\end{figure}
The Bayes factors between hypotheses $i$ and $j$, i.e. 
$BF_{i,j}=f(x=3\,|\,H_i)/f(x=3\,|\,H_j)$, are 
$BF_{2,1}=18$, $BF_{3,1}=25$ and $BF_{3,2}=1.4$.
The observation $x=3$ favors models 2 and 3, but the 
resulting probabilities depend on priors. {\it Assuming}
prior equiprobability among the three generators we get the
following posterior probabilities for the three models:
2.3\%, 41\% and 57\%. (In alternative, 
we could know that the extraction mechanism does not
choose the three generators at random with the same probability,
and the result would change.)

Instead, if we made an analysis based on p-value we would get 
that $H_1$ is ``excluded'' at a 99.87\% C.L. or at 99.7\% C.L.,
depending whether a one-tail or a two-tail test is done.
Essentially, the perception that $H_1$ could be the correct cause 
of $x=3$ is about 10-20 times smaller than that given by 
the Bayesian analysis. As far as the comparison between 
$H_2$ and $H_3$ is concerned,
 the p-value analysis is in practice inapplicable
(what would you do?)
and one says that both models describe about 
equally well the result, which is more or less what we 
get out of the Bayesian analysis. However, the latter
analysis gives some quantitative information: a slight hint
in favor of $H_3$, that could be properly combined
with many other small hints coming from other pieces of 
experimental information, and that, all together,
might allow us to finally arrive 
to select one of the  models.

\end{description}

\section{Conclusions}
The main message of this contribution is 
an invitation to review critically several concepts
and methods to which we are somehow accustomed.

Strict falsificationism is definitely na\"\i ve and 
its implementation via frequentistic hypothesis tests
is  logically seriously flawed. Such tests `often 
work' --- unfortunately I cannot not go through this point 
for lack of space and I refer to Section 10.8 of Ref.~\cite{BR} ---
if we want to use them to form a rough idea about whether it
 is worth investigating in alternative hypotheses 
that would describe the data better.
Stated in different words, there is nothing to reproach
--- and I admit I do it ---
calculating a $\chi^2$ variable to get a idea of the `distance'
between a model and the data.
What is not correct is to use the $\chi^2$, or any 
other test variable to quantitatively assess 
 our confidence on that model.

An alternative way of reasoning, based on probability theory
and then capable to quantify consistently
 our confidence in formal probabilistic terms, has been shortly outlined.  
I hope that, also with the help
of  the simple examples, the paper has been able to convey
some important points.
\begin{itemize}
\item
Bayes theorem provides a consistent way to learn from 
data both for probabilistic parametric inference and 
probabilistic model comparison.
\item
In order to perform a model comparison
at least two fully specified hypotheses are needed
[i.e. of which
we are able to evaluate, though roughly, the likelihood
$f(\mbox{data}\,|\,H_i(\mvec{\theta}))$, 
where $\mvec\theta$ are the model parameters].
\item
Scientific conclusions, i.e. how much we believe
in either hypothesis, must depend on priors
--- would you trust an `ad hoc' model tailored 
on the data you are going to use for the inference? 
\item 
If a hypothesis is hardly believable with respect to an
alternative hypothesis, then it is absolutely normal
that a stronger evidence in favor of it is needed, before we reverse
our preference.
\item
The Bayes factor can be considered an unbiased 
way to report how much the data alone `push' towards each 
hypothesis. 
\end{itemize}
An example of model comparison applied to real 
data,
in tune with the workshop themes and written also with
didactic intent, can be found in 
Ref.~\cite{ADD}.\footnote{Reference \cite{ADD} has to be taken
more for its methodological contents than
for the physical outcome (a tiny piece of evidence in favor
of the searched for signal), 
for in the meanwhile I have
become personally very sceptical about the experimental 
data on which the analysis was based,  
after having heard a couple of public talks by authors of those data
during 2004 (one in this workshop).}

Another important class of applications, not discussed in this paper, 
concerns parametric inference. 
Essentially, one starts from Eq.~(\ref{eq:Bayes_2}), and all the 
rest is `just math', including the extensions to several dimensions
and some `tricks' to get the computation done.
It can be easily shown that standard methods can be recovered
as approximated application of the Bayesian inference 
under {\it some well defined assumptions} 
that usually hold in routine applications.
I refer to Refs.~\cite{BR} and \cite{RPP}
for details concerning this point, as well as for other 
issues in Bayesian data analysis not discussed here, 
and a rich bibliography.

Finally, I would like to add some epistemological remarks. 
The first one concerns falsificationism, since after my conference talk
I have received quite some energetic reactions of colleagues 
who defended that principle.
From a probabilistic perspective, falsificationism is easily 
recovered if the likelihood vanishes, i.e.
 $f(\mbox{data}\,|\,H_i)=0$. 
However this condition is rarely met in the scientific practice, 
if we speak rigorously (zero is a very committing value!). 

I guess we just speak of falsificationism 
because that is what we have being taught is the `good thing', 
but without being aware of its implications. 
It seems to me we actually think in terms of something that should
better be named 
{\it testability}, that can be stated quite easily in the language 
of probabilistic inference. Given a hypothesis $H_i$, testability 
requires that the likelihood is positive in a region $Q$ of the space 
of the {\it achievable experimental outcomes} of an experiment $Exp$
[i.e. $f(\mvec{x}\, \mbox{in}\, Q\,|\,H_i,\,  Exp) \ne 0$] 
and is not trivially proportional to the likelihood of another hypotheses
[i.e. $f(\mvec{x}\, \mbox{in}\, Q\,|\,H_i, \, Exp) 
/f(\mvec{x}\, \mbox{in}\, Q\,|\,H_j,\, Exp) \ne k$].
These are in fact the conditions for a hypothesis to gain in credibility,
via Bayes theorem,
over the alternative hypotheses in the light of the 
expected experimental results.
The theory is definitively
falsified if the experimental
outcome falls on another region $Q^\prime$ such that 
$f(\mvec{x}\, \mbox{in}\, Q^\prime\,|\,H_i,\,  Exp) = 0$. 
Therefore, falsificationism is just a special case of the 
Bayesian inference. 

Anyway, if there is a topic in  which falsificationism
can be applied in a strict sense, this topic concerns 
the use of conventional statistical methods,
 as I wrote elsewhere\,\cite{BR}: 
``{\it I simply apply scientific methodology 
to statistical reasoning in the same way as we apply it
in Physics and in Science in general. 
If, for example, 
 experiments show that Parity is violated, we can be
disappointed, but we simply give up the 
principle of Parity Conservation,
at least in the kind of interactions in which it has been observed
that it does not hold. 
I do not understand why most of my colleagues do not behave in 
a similar way with the Maximum Likelihood principle, or with 
the `prescriptions' for building Confidence Intervals, both of which 
are known to produce absurd results.}''

The second epistemological remark  
concerns another presumed myth of scientists,
i.e. that ``{\it since Galileo an accepted base 
of scientific research is the
repeatability of experiments}.''\cite{Giunti}  
({\it ``This assumption justifies the Frequentistic definition 
of probability \ldots}'' --- continues the author.)
Clearly, according to this point of view, most 
things discussed in this workshop are 'not scientific'. 
Fortunately, it is presently rather well accepted 
(also by the author of Ref.~\cite{Giunti}, I understand)
that Science 
can be also based on a collection of individual facts that 
we cannot repeat at will, or that 
might happen naturally and beyond our control
(but there is still someone claiming fields like Geology, 
Evolutionary Biology and even Astrophysics are not Science!).
The relevant thing that allows us to build up 
a rational scientific
knowledge grounded on empirical observations
is that we are capable to relate, though in a stochastic way and 
with the usual  unavoidable uncertainties, our conjectures
to experimental observations, no matter if the phenomena occur 
spontaneously
or arise under well controlled experimental conditions.
In other words, we must be able to model, though  
approximately, the likelihoods that connect hypotheses to observations.
 This way of building the scientific edifice
is excellently expressed in the title of one of the volumes issued 
to celebrate the Centennial of the Carnegie Institute of 
Washington~\cite{Carnegie}.
This scientific building can be formally (and graphically) 
described by the so called `Bayesian networks' or 
`belief networks'~\cite{Causality}. If you have never heard these expressions, 
try to google them and you will discover a new world 
(and how behind we physicists are, mostly sticking to books 
and lecture notes that are too often copies of copies of obsolete books!).

{\sl It is a pleasure to thank the organizers
for the stimulating workshop in such a wonderful location, 
Paolo Agnoli and Dino Esposito for useful comments.}

\end{document}